\documentclass[sigconf]{acmart}

\AtBeginDocument{%
  \providecommand\BibTeX{{%
    \normalfont B\kern-0.5em{\scshape i\kern-0.25em b}\kern-0.8em\TeX}}}

\AtBeginDocument{%
 \providecommand\BibTeX{{%
    \normalfont B\kern-0.5em{\scshape i\kern-0.25em b}\kern-0.8em\TeX}}}
    
\usepackage{enumitem}
\usepackage{multirow}
\usepackage{makecell}
\usepackage{amsthm}
\usepackage{amsfonts}
\usepackage{url}
\usepackage{amsmath}
\DeclareMathOperator*{\argmax}{arg\,max}

\newcommand{\ie}{\emph{i.e., }}
\newcommand{\eg}{\emph{e.g., }}

\newcommand{\wrt}{\emph{w.r.t. }}

\newcommand{\aka}{\emph{aka. }}

\copyrightyear{2023}
\acmYear{2023}
\setcopyright{rightsretained}
\acmConference[KDD '23]{Proceedings of the 29th ACM SIGKDD Conference on Knowledge Discovery and Data Mining}{August 6--10, 2023}{Long Beach, CA, USA} \acmBooktitle{Proceedings of the 29th ACM SIGKDD Conference on Knowledge Discovery and Data Mining (KDD '23), August 6--10, 2023, Long Beach, CA, USA} \acmDOI{10.1145/3580305.3599285} \acmISBN{979-8-4007-0103-0/23/08}

\makeatletter
\gdef\@copyrightpermission{
  \begin{minipage}{0.3\columnwidth}
   \href{https://creativecommons.org/licenses/by/4.0/}{\includegraphics[width=0.90\textwidth]{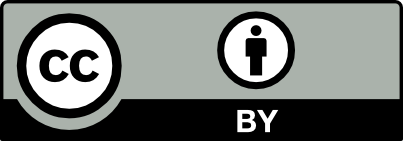}}
  \end{minipage}\hfill
  \begin{minipage}{0.7\columnwidth}
   \href{https://creativecommons.org/licenses/by/4.0/}{This work is licensed under a Creative Commons Attribution International 4.0 License.}
  \end{minipage}
  \vspace{5pt}
}
\makeatother

\begin{document}

\title{Context-aware Event Forecasting via Graph Disentanglement}

\author{Yunshan Ma}
\authornote{Both authors contributed equally to the paper.}
\affiliation{
    \institution{National University of Singapore}
    \country{}
}
\email{yunshan.ma@u.nus.edu}
 
\author{Chenchen Ye}
\authornotemark[1]
\country{}
\affiliation{
    \institution{National University of Singapore}
    \country{}
}
\email{chenchenye.ccye@gmail.com}

\author{Zijian Wu}
\country{}
\affiliation{
    \institution{National University of Singapore}
    \country{}
}
\email{zijian.wu@u.nus.edu}

\author{Xiang Wang}
\authornote{Corresponding author. Xiang Wang is also affiliated with Institute of Artificial Intelligence, Institute of Dataspace, Hefei Comprehensive National Science Center.}
\affiliation{
    \institution{University of Science and Technology of China}
    \country{}
}
\email{xiangwang1223@gmail.com}

\author{Yixin Cao}
\affiliation{
    \institution{Singapore Management University}
    \country{}
}
\email{caoyixin2011@gmail.com}

\author{Tat-Seng Chua}
\affiliation{
    \institution{National University of Singapore}
    \country{}
}
\email{dcscts@nus.edu.sg}

\begin{abstract}
Event forecasting has been a demanding and challenging task throughout the entire human history. It plays a pivotal role in crisis alarming and disaster prevention in various aspects of the whole society. The task of event forecasting aims to model the relational and temporal patterns based on historical events and makes forecasting to what will happen in the future. Most existing studies on event forecasting formulate it as a problem of link prediction on temporal event graphs. However, such pure structured formulation suffers from two main limitations: 1) most events fall into general and high-level types in the event ontology, and therefore they tend to be coarse-grained and offers little utility which inevitably harms the forecasting accuracy; and 2) the events defined by a fixed ontology are unable to retain the out-of-ontology contextual information.

To address these limitations, we propose a novel task of context-aware event forecasting which incorporates auxiliary contextual information. First, the categorical context provides supplementary fine-grained information to the coarse-grained events. Second and more importantly, the context provides additional information towards specific situation and condition, which is crucial or even determinant to what will happen next. However, it is challenging to properly integrate context into the event forecasting framework, considering the complex patterns in the multi-context scenario. Towards this end, we design a novel framework named \textbf{Se}paration and \textbf{Co}llaboration \textbf{G}raph \textbf{D}isentanglement (short as \textbf{SeCoGD}) for context-aware event forecasting. In the separation stage, we leverage the context as a prior guidance to disentangle the event graph into multiple sub-graphs, followed by a context-specific module to model the relational-temporal patterns within each context. In the collaboration stage, we design a cross-context module to retain the collaborative associations among multiple contexts.
Since there is no available dataset for this novel task, we construct three large-scale datasets based on GDELT. Experimental results demonstrate that our model outperforms a list of SOTA methods. The dataset and code are released via \url{https://github.com/yecchen/SeCoGD}.
\end{abstract}




\maketitle

\renewcommand{\shortauthors}{Yunshan Ma and Chenchen Ye et al.}

\section{Introduction} \label{sec:introduction}

\begin{figure}
    \centering
    \includegraphics[width = 0.95\linewidth]{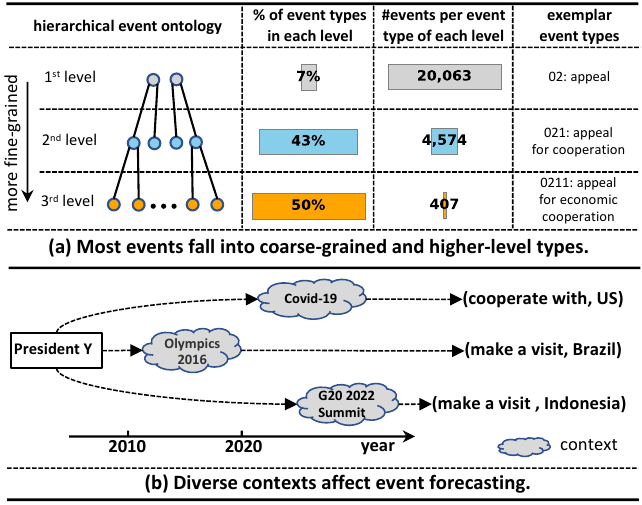}
    %\vspace{-5pt}
    \caption{The motivation of context-aware event forecasting. (a) Most current events fall in the coarse-grained and higher level types of the ontology, while more informative fine-grained events are fewer. (b) Out-of-ontology and diverse contexts affect events. Context can provide more fine-grained information to enhance the event forecasting performance.}
    \label{fig:motivation}
    \vspace{-15pt}
\end{figure}

Event forecasting~\cite{survey-event-forecasting} is one of the long-standing and challenging tasks, including forecasting of pandemic outbreak~\cite{disease}, civil unrest~\cite{icews}, international conflicts~\cite{gdelt}, \textit{etc}. Accurately predicting such vital events enables people to prepare in advance to prevent catastrophic results or minimize potential influence. Automatic event forecasting targets at modeling the rich relational and temporal patterns endowed by events observed in history, thus making accurate forecasting to events in the future. The development of data science and artificial intelligence endows human with stronger capability for automatic event forecasting, which has garnered more and more attention in recent years.

One of the prominent formulations for event forecasting is to define an event as a quadruple, \ie $(s, r, o, t)$, where $s$, $r$, $o$, and $t$ refer to subject, relation (event type)~\footnote{Relation and event type are the same in this work.}, object, and timestamp, respectively. At each timestamp, all the quadruples form an event graph. Given a query of $(s, r, t+1)$ in the future and the list of historical event graphs until $t$, we aim to predict the missing object. Based on such a structured formulation, a plethora of works have been emerging in recent years. They have applied structured relational and temporal information (\eg RE-NET~\cite{RENET}, RE-GCN~\cite{REGCN}), time intervals (\eg EvoKG~\cite{EvoKG}), and texts from ontology and news articles (\eg Glean~\cite{Glean} and CMF~\cite{CMF}), \textit{etc.} for event forecasting.

%RE-NET~\cite{RENET} and RE-GCN~\cite{REGCN} are pioneering works that make use of graph neural network (GNN) to retain the relations among concurrent events and employ recurrent neural network (RNN) to capture the temporal evolution of events. EvoKG~\cite{EvoKG} emphasizes the role of time and innovatively integrates both time and event prediction into a unified framework. In addition to the structured data, Glean~\cite{Glean} and CMF~\cite{CMF} resort to hierarchical and heterogeneous data to enhance the event forecasting accuracy as well as generate explanations along with the prediction.

Albeit the remarkable achievements of current works~\cite{RENET,REGCN,Glean,CMF,TKG-survey}, they still suffer from the following limitations. First, existing structured events tend to be classified as high-level general events, while more specific and informative events are few. As shown in Figure~\ref{fig:motivation}(a), for the well-known GDELT~\cite{gdelt} dataset, while the hierarchical event type ontology~\cite{CAMEO} defines a large number of fine-grained event types, most actual events were being classified into event types in the higher levels of event ontology and fine-grained events are fewer. Consequently, the expressiveness of events is severely restricted, resulting in less utility in practical scenarios. Second, events follow the types in the predefined ontology, which is usually fixed due to the difficulty of construction. For example in political event forecasting, 
%the famous WEIS~\cite{WEIS} ontology was curated during Cold War and had been used for four decades~\cite{CAMEO} until the curation of 
the well-known CAMEO~\cite{CAMEO} ontology costs ten years to be finalized. It is difficult to update the ontology timely, thus newly emerging out-of-ontology information is unable to be covered by the outdated ontology~\footnote{Both out-of-ontology and outdated ontology refer to the same problem in this work.}. Worse still, events are greatly influenced by out-of-ontology \textit{contextual} information, such as the situation and circumstance. As the example shown in Figure~\ref{fig:motivation}(b), given diverse contexts, the entity \textit{President Y} performs distinct roles and actions \wrt the various countries of the world. Such diverse contexts that provide clues for certain situations, are crucial or even determinant to event forecasting, and they cannot be adequately modeled solely based on the event ontology.

To address these limitations, we introduce \textit{context} into existing event representation as supplementary information and define a novel task named \textit{context-aware event forecasting}. We associate each event with a categorical context, elaborating the event's occurrence situation or condition. Then each event is extended from a quadruple to a quintuple, \ie $(s, r, o, t, c)$, where $c$ denotes the context. The incorporation of context brings multiple benefits to event forecasting. First, it endows more fine-grained information to each event, thus making the coarse-grained events more specific and expressive. Second, the flexibly defined context is able to offer crucial information about the circumstances or backgrounds of the events, narrowing down the potential forecasting space. As the example shown in Figure~\ref{fig:motivation}(b), given different contexts of \textit{Olympics 2016} or \textit{G20 2022 Summit}, the target countries that \textit{President Y} would make a visit to will be different. 
%Brazil to attend the opening ceremony. Meanwhile, in the context of , \textit{President Y} would make a visit to Indonesia to attend the summit in Bali.
%The context can be flexibly defined and labelled to existing events. For example, a context can be: 1) a specific condition, such as Israel-Arabs or Russia-Ukraine issues, which is not an event type in current ontology; 2) a newly-emerging event type, such as Covid-19 pandemic, which has not yet been covered by current ontology; 3) an existing event type, such as economical or political, which may not be well extracted from the raw data.
%During inference, people can pair a query with a specific context as explicit instruction, thus enabling forecasting in targeted condition. 
Despite various merits, integrating context into the problem of event forecasting poses new challenges to existing methods. %First, most current approaches cannot handle the supplementary information of context, and it is challenging to design an effective framework to simultaneously model both context-aware and original relational-temporal patterns.
%First, the context preserves fine-grained supplementary information, how to leverage such information thus enhance event forecasting is not trivial.
First, one entity under different contexts may trigger distinctive events, and how to capture the relational and temporal patterns given a certain context is not trivial. Second, events from different contexts are also correlated with each other, and how to delicately model the collaborative associations among contexts is vital to accurate event forecasting.

To tackle the above challenges,  we borrow the idea from graph disentanglement representation learning and propose a general framework \textbf{SeCoGD} (\textbf{Se}paration and \textbf{Co}llaboration \textbf{G}raph \textbf{D}isenta-nglement), for context-aware event forecasting. 
%First, to capture the fine-grained information, we borrow the idea from graph disentanglement and utilize the context as a prior guidance to separate the event graphs into multiple sub-graphs. Second, 
It consists of two stages: \textit{separation} and \textit{collaboration}. First, in the separation stage, we utilize the context as a prior guidance to separate the event graphs into multiple sub-graphs. Then, we resort to established relational-temporal models, such as RE-GCN~\cite{REGCN}, to capture the context-specific patterns within each sub-graph. Second, in the collaboration stage, we construct hypergraphs among the disentangled embeddings and leverage GNN to learn the collaborative associations among contexts. Different from current graph disentanglement methods that just focus on how to separate the graph, our framework considers both separation and collaboration due to the prior guidance of context for the separation. Moreover, our method is a general framework, of which the key modules can be replaced by alternative designs. At the same time, considering that there is no available dataset that has context information, we build three new datasets based on GDELT. Extensive experiments demonstrate that our proposed framework outperforms various SOTA methods. The main contributions of our work are summarized as:

\begin{itemize}[leftmargin=*]
    \item We propose to introduce the categorical contextual information into the structured event forecasting problem.
    \item To tackle the new task, we build a novel framework SeCoGD, and the two-stage design of separation and collaboration is effective in capturing the complex patterns in the multi-context scenario.
    \item We build three datasets based on GDELT to facilitate current and future studies for context-aware event forecasting. Our method significantly outperforms SOTA methods on the three datasets.
\end{itemize}

% change the challenges to map it with the two stages of separation and collaboration
% summarize the review part to highlight the key part: the information current methods used
% illustrate what is event graph in the review paragraph
% give an example in Figure 1(a) about what is high-level event types; put the capture on the bottom of the figure
% ethics claim in the footnote

% select another running example instead of the sensitive China vs US disputes, and most Singaporeans do not think in this way though..
% give a high-level definition of the events we are going to forecast
\section{Preliminary} \label{sec:problem_formulation}
We first give a formal definition for the task of context-aware event forecasting. Then, we introduce the newly constructed datasets.

\subsection{Problem Formulation} \label{subsec:problem_formulation}
We first present the problem formulation of conventional event forecasting, which does not consider the contextual information. Then we present how to introduce the context and formulate the new task of context-aware event forecasting. 

\textbf{Conventional Event Forecasting.} We define an event as a quadruple $(s, r, o, t)$, where $s \in \mathcal{E}$, $r \in \mathcal{R}$, and $o \in \mathcal{E}$ corresponds to subject entity, relation, and object entity, respectively; $t$ is the timestamp when this event happens; $\mathcal{E}$ and $\mathcal{R}$ are the entity and relation set, respectively. All the quadruples in the same timestamp $t$ form an event graph, denoted as $G_t=\{(s_n, r_n, o_n, t)\}_{n=1}^{N}$, where $(s_n, r_n, o_n, t)$ is the $n$-th event, and $N$ is the number of events at timestamp $t$. Given the historical event graphs at and before time $t$, denoted as $\mathbf{G}_{\le t}=\{G_1, G_2, \cdots, G_{t}\}$, and a query, denoted as $(s, r, t+1)$, we aim to predict the object $o$.

\textbf{Context-aware Event Forecasting.} We define context $c \in \mathcal{C}$ as a categorical value denoting certain situations or conditions shared by a group of events, where $\mathcal{C}=\{c_1, c_2, \cdots, c_K\}$ is the set of contexts and $K$ is the number of a few contexts. In practice, for historical events, the context can be obtained from human annotation, crowd-sourcing tags, or automatic information extraction systems. We assign a context $c$ to each event, thus extending its quadruple representation into a quintuple representation, denoted as $(s, r, o, t, c)$. Correspondingly, the event graph at each timestamp $t$ will be extended as $G_t=\{(s_n, r_n, o_n, t, c_n)\}_{n=1}^{N}$, where $c_n$ is the context of the $n$-th event. Given the historical event graphs $\mathbf{G}_{\le t}$, a query $(s, r, t+1)$ and a specified context $c$ in which the query event is supposed to be, we target at predicting the object $o$. Please note that specifying the \textbf{categorical} context during inference will not leak information about the predicted object. For example, given the context of Covid-19, the query "which country that President Y will cooperate with" will not be leaked to the model. And we assume that it is not difficult for human to provide such contextual information for a certain event he/she wants to predict. 

%we have ground-truth contexts for events during inference. Because according to our definition of context, it is a categorical value specified by human, and we assume it is not difficult for human to provide such contextual information for a certain event he/she wants to predict.

\subsection{Dataset Construction} \label{subsec:dataset_construction}
Existing datasets for event forecasting are different cropped versions of GDELT~\cite{gdelt} and ICEWS~\cite{icews}. For example, among the datasets used by current works~\cite{RENET,REGCN,TANGO,xERTE,HiSMatch}, ICEWS14, ICEWS18, ICEWS05-15 include events in the ICEWS dataset of year 2014, 2018, and 2005-2015, respectively; and GDELT covers January 2018 of the original GDELT dataset. However, all of these versions only use the existing quadruple data while overlooking the context information.

To facilitate the study of context-aware event forecasting, we build three benchmark datasets based on the GDELT dataset~\cite{gdelt}, which provides the original news article URLs of the extracted events. Following previous works~\cite{Glean,CMF}, we crop three subsets of GDELT according to the regions of the events, \ie Egypt (EG), Iran (IR), and Israel (IS), spanning from February 2015 to March 2022. According to a previous systematical study~\cite{compare-gdelt-icews}, the structured events extracted by GDELT have high recall while low precision, which means there are many false positive events. Such noise could be caused by the event extraction system used in GDELT or the low quality of original articles. Since the GDELT event extraction system is unavailable to the public, we aim to remove low-quality articles to eliminate these noises by the following data preprocessing steps. 

First, we keep the event with a valid URL. Second, we sort the domain names of the URLs, which correspond to different news agencies. 
In total, there are around 20K domain names, and the top 69 cover 40\% of the events. After checking these top domain names, we confirm that their news articles are of higher quality and reliability. Therefore, we remove the remaining 60\% of events that are published in long-tailed domain names, which are usually from less influential agencies or personal blogs and are likely to be of low quality or even fake. Third, even though the interval of two consecutive timestamps in the original GDELT data is 15 minutes, it is unnecessary to have such precise timestamps for political events. Following ICEWS, we take the one-day time interval and collapse the 15 minutes-level timestamps of events on the same day to the day-level timestamp. Finally, we obtain the datasets and split them into training/validation/testing with a ratio of 8/1/1 over the timeline. Due to the severe popularity bias of the dataset, several most frequent entities in the validation and testing set are masked. The statistics are shown in Table~\ref{tab:dataset}.

\begin{table}[t]
\begin{center}
%\vspace{-0.1in}
\caption{Dataset Statistics.}
\label{tab:dataset}
\vspace{-0.1in}
\resizebox{0.49\textwidth}{!}{
    \begin{tabular}{cccccccc}
        \hline
         & $\lvert \mathcal{V} \rvert$ & $\lvert \mathcal{E} \rvert$ & \#urls & \#days & \#train & \#valid & \#test \\
        \hline
        EG & 2,594 & 225 & 96,081 & 2,584 & 377,430 & 36,588 & 28,644 \\
        IR & 2,988 & 236 & 223,616 & 2,584 & 973,752 & 69,827 & 76,239 \\
        IS & 3,456 & 238 & 345,611 & 2,584 & 1,430,389 & 171,518 & 156,695 \\
        \hline
    \end{tabular}
}
\end{center}
\vspace{-0.15in}
\end{table}

Since there is no context label in the original GDELT dataset, we leverage the textual content and topic model (\ie LDA~\cite{LDA}) as a \textit{proxy} to generate contexts. In particular, we use an LDA model to extract the topic distribution of each article, where the topic with the highest weight is treated as the context of an article. An event is assigned with its corresponding article's context. More studies about the topic models as well as alternative context generation approaches are presented in Section~\ref{subsec:context_study}. To be noted, during inference, people can provide certain context for the query to be predicted.
\section{Methodology} \label{sec:methodology}
 
\begin{figure*}
    \centering
    \includegraphics[width = 0.95\linewidth]{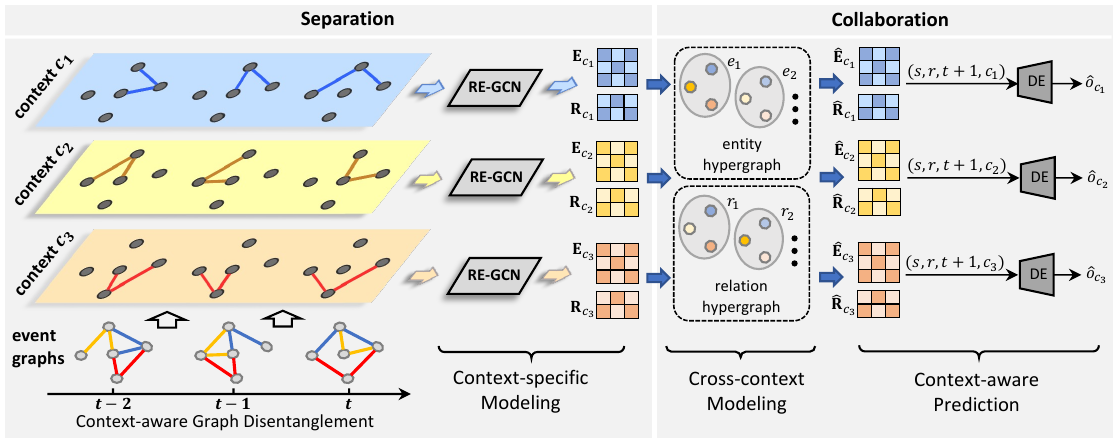}
    \vspace{-0.1in}
    \caption{The overall framework of SeCoGD consists of two stages: separation and collaboration. The separation stage includes the context-aware graph disentanglement and context-specific modeling modules, and the collaboration stage comprises the cross-context modeling and context-aware prediction modules.}
    \vspace{-0.15in}
    \label{fig:framework}
\end{figure*}

To solve the new problem of context-aware event forecasting, we propose a novel framework \textbf{Se}paration and \textbf{Co}llaboration \textbf{G}raph \textbf{D}isentanglement (\textbf{SeCoGD}), as shown in Figure~\ref{fig:framework}. It consists of two stages: the separation stage and the collaboration stage.

\subsection{Separation}
In the separation stage, we first use the context as a prior guidance to disentangle the event graph into multiple sub-graphs. Then we devise a context-specific modeling module to capture the relational and temporal patterns within each context.

\subsubsection{Context-aware Graph Disentanglement}
Generally, events in the same context exhibit similar or correlated patterns, while events in different contexts demonstrate distinctive characteristics. Current works~\cite{RENET,REGCN,EvoKG,CMF} connect all the quadruples at the same timestamp as a unified event graph and learn a single embedding for each entity and relation via GNN models. However, such unified entity and relation embeddings are highly entangled \wrt diverse context~\cite{DisenGCN}, failing to capture the context-specific patterns. 

Inspired by recent progress in disentangled representation learning~\cite{DisenGCN,DisenHAN,DisenKGAT}, we seek graph disentanglement for context-aware event forecasting. Most existing works solely rely on the inherent structural information for graph disentanglement. For example, MaridVAE~\cite{MacridVAE} and DGCF~\cite{DGCF} utilize the user-item interactions to learn disentangled representations for different intents; DisenKGAT~\cite{DisenKGAT} tackles the heterogeneous knowledge graph and disentangles the entity embedding with respect to different topics and clusters. Nonetheless, these methods are incapable of disentangling event graphs since the events are too coarse-grained, and pure structural information is unable to well disentangle the graph. 

We employ the context as a prior guidance to disentangle the event graph. Formally, given $K$ contexts, we separate the original entangled event graph $G_t$ into $K$ sub-graphs $\{G^{c_1}_t, \cdots, G^{c_k}_t, \cdots, G^{c_K}_t\}$, where each sub-graph $G^{c_k}_t$ contains all the events within context $c_k$, denoted as $G^{c_k}_t = \{(s_n, r_n, o_n, c_k, t)\}_{n=1}^{N^{c_k}_t}$, where $N^{c_k}_t$ is the number of events in timestamp $t$ within context $c_k$. Note that we make use of the external prior knowledge, \ie the context, to disentangle the original graph. Meanwhile, previous works exploit end2end solutions, which either adopt attention mechanism~\cite{DisenGCN} or incorporate auxiliary distance regularizer~\cite{DGCF} to directly learn disentangled representations solely relying on the graph data. This prior-guided disentanglement is better than the end2end solutions in separating the graph due to the incorporation of external knowledge. Based on the disentangled event graph, the core of the model is to model the patterns within each context and across multiple contexts. 

%three key components, \ie context-specific modeling, cross-context modeling, and context-aware prediction, are designed to perform representation learning and event forecasting. We present the three modules in the subsequent sections.

\subsubsection{Context-specific Modeling} \label{subsec:separation}
Each separated graph preserves distinct characteristics for both concurrent relations and evolving patterns under its corresponding context. Towards this end, we build a context-specific modeling module for each context. Given a list of historical event graphs $G^c_{\le t}$ under a certain context $c$, the context-specific modeling module aims to learn entity and relation representations. We inherit the design of RE-GCN~\cite{REGCN} to build this context-specific modeling module, which encompasses two parts: concurrent events modeling and temporal event modeling.  

\textbf{Concurrent Event Modeling} is curated to model the relationship among events occurring in the same timestamp. We make use of RGCN~\cite{RGCN}, which is capable of modeling multi-relation graphs, as the graph kernel to learn the entity representation. At timestamp $t$ under context $c$, for each layer $l$ of the graph propagation, the message obtained by each object $o$ is $\mathbf{e}^{l}_{o,t,c} \in \mathbb{R}^{d}$, defined as:
\begin{equation} \label{eq_1}
    \mathbf{e}^{l}_{o} = f \Biggl( \frac{1}{\vert \mathcal{E}_{o} \rvert} \sum_{(s,r) \in \mathcal{E}_{o}}{\mathbf{W}^l_1(\mathbf{e}^{l-1}_{s} + \mathbf{r}) + \mathbf{W}^{l}_{2} \mathbf{e}^{l-1}_{o}} \Biggr), 
\end{equation}
where $d$ is the dimensionality of the message, $\mathcal{E}_{o}$ stands for all the events of which $o$ is the object, $\mathbf{W}^l_1, \mathbf{W}^l_2 \in \mathbb{R}^{d \times d}$ are the parameters of the convolutional kernel in layer $l$, and $f(\cdot)$ is the activation function which we use RReLU. To be noted, $\mathbf{e}^{l}_{o}, \mathcal{E}_{o}, \mathbf{e}^{l-1}_{s}, \mathbf{r}, \mathbf{e}^{l-1}_{o}$ all stand for their corresponding representations in time $t$ of context $c$, where we omit the subscript $(t, c)$ for simplicity. After performing multi-layer message passing, we aggregate the messages obtained from multi-layer propagation and yield the entity representation at time $t$ under context $c$, defined as:
\begin{equation} \label{eq_2}
    \mathbf{e}_{o,t,c} = \sum^L_{l=0}{\mathbf{e}^l_{o,t,c}},
\end{equation}
where $\mathbf{e}^0_{o,t,c}=\mathbf{e}^0_{o,c} \in \mathbb{R}^{d}$ is randomly initialized for each entity $o$ under each context $c$. And the representations for all entities at timestamp $t$ within context $c$ are denoted as $\mathbf{E}^{\prime}_{t,c} \in \mathbb{R}^{| \mathcal{E} | \times d}$.

\textbf{Temporal Pattern Modeling} is designed to capture the temporal evolution of entities and relations. Following the previous study~\cite{REGCN}, we devise a learnable gate mechanism to reserve the entities' evolving patterns. It is formally defined as:
\begin{equation} \label{eq_3}
    \mathbf{E}_{t,c} = \mathbf{U}_{t,c} \mathbf{E}^{\prime}_{t,c} + (1 - \mathbf{U}_{t,c}) \mathbf{E}_{t-1,c},
\end{equation}
where $\mathbf{U}_{t,c} \in \mathbb{R}^{d \times d}$ is the learnable gate, which is calculated by a nonlinear transformation:
\begin{equation} \label{eq_4}
    \mathbf{U}_{t,c} = \sigma (\mathbf{W}_4 \mathbf{E}_{t-1,c} + \mathbf{b}),
\end{equation}
where $\sigma(\cdot)$ is the sigmoid activation function, and $\mathbf{W}_4$ and $\mathbf{b}$ are trainable parameters for the gate. For efficiency during implementation, we take the recent $D$ steps of historical graphs to capture the temporal evolving patterns, following the typical TKG (Temporal Knowledge Graph) solutions~\cite{TKG-survey}. Then, the entity embeddings in the last step preserve all the context-aware relational and temporal patterns, and we denote them as $\mathbf{E}_{c}$.

For the relation representation, we concatenate its embedding and associated entities, thus each relation embedding is updated as:
\begin{equation} \label{eq_5}
    \mathbf{r}^{\prime}_{t,c} = \left[ \mathbf{r}_c; \frac{1}{\lvert \mathcal{V}_{r,t,c} \rvert} \sum_{v \in \mathcal{V}_{r,t,c}}{\mathbf{e}_{v,t,c}} \right],
\end{equation}
where $\mathcal{V}_{r,t,c}$ is the set of entities that connect to the relation $r$, $\mathbf{e}_{v,t,c} \in \mathbb{R}^{d}$ is the representation of entity $v$ in $\mathbf{E}_{t,c}$, and $\left[;\right]$ is the concatenation operation. Then a GRU is applied to deduce the temporal relation representation $\mathbf{r}_{t,c}$, calculated by:  

\begin{equation} \label{eq_6}
    \mathbf{r}_{t,c} = \text{GRU} \bigl( \mathbf{r}_{t-1,c}, \mathbf{r}^{\prime}_{t,c} \bigr).
\end{equation}
And all the relations' representations at time $t$ for context $c$ are defined as $\mathbf{R}_{t,c} \in \mathbb{R}^{| \mathcal{R} | \times d}$. Going through $D$ steps of recurrent units, we obtain all the relations' representations that retain relational and temporal information conditioned on context $c$, denoted as $\mathbf{R}_{c}$. 

\subsection{Collaboration} \label{subsec:collaboration}
In the collaboration stage, we leverage hypergraphs to model the cross-context collaborative associations. Then we perform context-aware prediction and optimization.

\subsubsection{Cross-context Modeling}
Even though the same entity demonstrates different characteristics in various contexts, these contexts are not independent but correlated with each other. 
%For example, given the three exemplar contexts of China-US Trade War, Covid-19 Pandemic, and Russia-Ukraine War, some policies of a certain country must consider two or three contexts simultaneously, such as taking a position among the three major countries of US, China, and Russia. 
For example, given the contexts of \textit{Covid-19 Pandemic} and \textit{Russia-Ukraine War}, many countries must consider them simultaneously to make economic policies, in order to minimize the influence on their economy as well as social stability.
To this end, capturing such correlation is crucial for some events that are affected by multiple contexts. Furthermore, after disentangling the event graph into multiple contexts, each sub-graph will be sparser than the original unified graph. Some entities that do not have sufficient occurrence in a certain context will not be well-trained for accurate forecasting. For such few-shot entities and relations, transferring knowledge from other contexts that have sufficient training data is a promising solution. 

Based on the above motivations, we devise a collaboration module to model the collaborative effects among multiple contexts, aiming to achieve potential knowledge transfer for sparse entities. It is worth mentioning that we do not have supervised information to quantify the correlations among contexts, thus, we are unable to explicitly model the collaborative effects. Considering this, we resort to hypergraph to model the latent collaborations. Concretely, for each entity $v$, we construct a hypergraph among its sub-embeddings in different contexts, where the nodes are the separated embeddings of all entities (relations) in different contexts and every hyper-edge connects the separated embeddings of the same entity (relation). Then we leverage a multi-layer LightGCN~\cite{LightGCN} to propagate over every hypergraph, and $\mathbf{\hat{e}}^p_{v,c} \in \mathbb{R}^{d}$ is the $p$-th layer propagated information to node $v$ under context $c$, obtained by:
\begin{equation} \label{eq_7} 
    \mathbf{\hat{e}}^p_{v,c} = \frac{1}{|\mathcal{C}_{v}|-1} \sum_{i \in \mathcal{C}_{v} \backslash \{c\}}{\mathbf{\hat{e}}^{p-1}_{v,i}},
\end{equation}
where $\mathcal{C}_{v}$ are all the contexts that the entity $v$ has been in.
After $P$ layers of propagation, we aggregate each layer's embedding and yield the final entity representation:
\begin{equation} \label{eq_8}
    \mathbf{\hat{e}}_{v,c} = \sum^P_{p=0}{\mathbf{\hat{e}}^{p}_{v,c}},
\end{equation}
where $\mathbf{\hat{e}}^0_{v,c}$ is the representation of entity $v$ in $\mathbf{E}_c$. After the hypergraph propagation, all entities are represented as $\mathbf{\hat{E}}_{c}$.

Analogous to entities, relations' representations in different contexts are also totally isolated during the context-specific modeling. Thereby, for each relation, we also build a hypergraph and take advantage of a multi-layer LightGCN kernel to capture the collaborative associations among different contexts, defined as:
\begin{equation} \label{eq_9} 
    \mathbf{\hat{r}}^p_{x,c} = \frac{1}{|\mathcal{C}_{x}|-1} \sum_{i \in \mathcal{C}_{x} \backslash \{c\}}{\mathbf{\hat{r}}^{p-1}_{x,i}},
\end{equation}
where $\mathbf{\hat{r}}^p_{x,c} \in \mathbb{R}^{d}$ is information propagated to relation $x$ in layer $p$, and $\mathcal{C}_{x}$ is the set of contexts that relation $x$ has been in over the historical observations. With $P$ layers of graph propagation, we aggregate multiple layers' representations and obtain the final relation embedding $\mathbf{\hat{r}}_{x,c}$, formally written as: $\mathbf{\hat{r}}_{x,c} = \sum^P_{p=0}{\mathbf{\hat{r}}^{p}_{x,c}}$,
%\begin{equation} \label{eq_10}
%    \mathbf{\hat{r}}_{x,c} = \sum^P_{p=0}{\mathbf{\hat{r}}^{p}_{x,c}},
%\end{equation}
where $\mathbf{\hat{r}}^{0}_{x,c}$ is relation $x$'s embedding in $\mathbf{R}_c$.

\subsubsection{Context-aware Prediction and Optimization} \label{subsec:event_forecasting}
With the context-specific and cross-context modeling modules, we learn the entity and relation representations that not only capture context-aware characteristics but also preserve transferred knowledge from other contexts. Following the established approach to event forecasting~\cite{REGCN,HiSMatch}, we devise a decoder based on ConvTransE~\cite{ConvTransE}. In particular, given a query quadruple $(s, r, t, c)$, we first use a ConvTransE to produce the query's representation, then score the candidate entities $\mathcal{E}$ via inner-product between the query and candidate representations. Formally, we calculate the prediction scores for all candidate entities given the query $(s,r)$ at time $t+1$ under context $c$ as follows: 
\begin{equation} \label{eq_11}
    \mathbf{\hat{p}}(\mathcal{E}|s,r,c,G_{\le t}) = \text{softmax} \bigl( \mathbf{\hat{E}}_{c} \text{ConvTransE}(\mathbf{\hat{e}}_{s,c}, \mathbf{\hat{r}}_{c}) \bigr),
\end{equation}
where $\text{softmax}(\cdot)$ is the softmax function, $\text{ConvTransE}(\cdot)$ is the ConvTransE decoder, and $\mathbf{\hat{e}}_{s,c}$ and $\mathbf{\hat{r}}_{c}$ are the representations for $s$ and $r$, respectively. The predicted object is presented as: 
\begin{equation} \label{eq_12}
    \hat{o}_{(s,r,t+1,c)} = \argmax_{\mathcal{E}}{\hat{p}(\mathcal{E}|s,r,c,G_{\le t})}.
\end{equation}

We employ cross-entropy loss to optimize the whole framework in an end-to-end fashion, and the loss is defined as:
\begin{equation} \label{eq_13}
    \mathcal{L} = \sum_{t=0}^{T-1} \sum_{c \in \mathcal{C}} \sum_{(s,r) \in G^c_{t+1}} { \mathbf{y}_{(s,r,t+1,c)} \text{log} \mathbf{\hat{p}}(\mathcal{E}|s,r,c,G_{\le t})},
\end{equation}
where $T$ is the total number of timestamps in the training set, and $\mathbf{y}_{(s,r,t+1,c)}$ is the one-hot representation of ground-truth object $o$.

\subsection{Discussion} \label{subsec:discussion}
To further highlight the key contributions of this work, we discuss the generalization capability of SeCoGD, as well as the rationale behind separation and collaboration.  
\subsubsection{Generalization Capability} 
We argue that our method SeCoGD is a general framework instead of a specific model. The key contribution of SeCoGD lies in two aspects: 1) it makes use of context as a prior guidance to disentangle the event graph; and 2) it proposes a novel graph disentanglement idea under prior-guided disentanglement, that is to model the collaborative association among the disentangled representations. First, the context can be flexibly defined according to various application scenarios. For example, the tag of the news article that an event belongs to can be used as its context. Alternatively, similar to our solution of the latent topic model, various automatic text clustering algorithms, such as K-means or GMM (Gaussian Mixture Model), are plausible to identify the latent contexts of events. Second, each component of SeCoGD has various alternatives. For example, RE-NET~\cite{RENET} can be used to replace the RE-GCN module for context-specific modeling, hypergraph can be replaced by some regularizers (\ie L2 distance) that pull closer the disentangled representations, and multiple modules~\cite{ConvE,DistMult,RotatE} could be manipulated as the decoder. 

\subsubsection{Separation and/or Collaboration}
Our work strengthens that it is crucial to incorporate the collaboration stage on top of the separation stage. However, most previous works on graph disentanglement solely focus on the separation part. For example, they either leverage regularization terms to maximize the mutual information~\cite{DGCF} among multiple chunked representations or use attention mechanism~\cite{DisenGCN} to make different disentangled representations attend on various sub-graphs. We assume that such contradictory modeling philosophy roots in two reasons. First, we have prior knowledge as the guidance for the disentanglement, therefore, we do not need any heuristically manipulated disentanglement strategies, such as mutual information maximization or attention. Second and more importantly, we believe that the crux of an effective graph disentanglement model lies in a good balance of separation and collaboration. Current works are built upon a unified graph model, which is highly intertwined. Thereby, a separation module is necessary to eliminate the entanglement. Meanwhile, we disentangle event graphs by the prior contextual information, where the sub-graphs are well or even over separated, thus a collaboration module is required to rectify the separation.
\section{Experiments} \label{sec:experiment}
We aim to answer the following research questions:

\begin{itemize}[leftmargin=*]
    \item \textbf{RQ1: } Does our framework outperform the SOTA methods?
    \item \textbf{RQ2: } Are the design of two stages, \ie separation and collaboration, effective in terms of event forecasting?
    \item \textbf{RQ3: } How does the context affect event forecasting?
\end{itemize}

\subsection{Experimental Settings}
We conduct experiments on the three datasets that we constructed, \ie EG, IR, and IS. The construction and statistics of the datasets can be found in Section~\ref{subsec:dataset_construction}. Following previous settings~\cite{REGCN}, we use Mean Reciprocal Rank (MRR) and HIT@\{1, 3, 10\} as the evaluation metrics. We use MRR to select the best model based on the validation set and record its corresponding performance on the testing set.

\subsubsection{Compared Methods}
Since current works have never studied the newly proposed problem of context-aware event forecasting on temporal event graph data, we select several strands of the most relevant works to compare with our proposed method. 

\begin{itemize}[leftmargin=*]
    \item \textbf{Static KG completion methods} treat event forecasting as a link prediction task on the static event graph. We select the following representative methods: \textbf{DistMult}~\cite{DistMult}, \textbf{ConvE}~\cite{ConvE}, \textbf{ConvTransE}~\cite{ConvTransE}, \textbf{RotatE}~\cite{RotatE}, and \textbf{RGCN}~\cite{RGCN}. 
    
    \item \textbf{Temporal KG forecasting methods} are designed for temporal event forecasting. These methods consider both relational and temporal information for link prediction in the next timestamp. We consider the following SOTA methods: \textbf{TANGO}~\cite{TANGO}, \textbf{RE-NET}~\cite{RENET}, \textbf{RE-GCN}~\cite{REGCN}, \textbf{EvoKG}~\cite{EvoKG}, and \textbf{HiSMatch}~\cite{HiSMatch}. 
    
    \item \textbf{Temporal event forecasting methods with texts} incorporate textual information into the event forecasting model, while the standard temporal KG forecasting methods only use structural information. In particular, we implement two versions of a representative method: 1) $\textbf{CMF}_{ont}$~\cite{CMF}, by faithfully following the settings of the original work and incorporating the event textual description defined in the CAMEO~\cite{CAMEO} \textit{ontology} into the structural event forecasting model. In addition, we also re-implemented 2) $\textbf{CMF}_{art}$~\cite{CMF}, which differs from $\textbf{CMF}_{ont}$ by using the original \textit{article} embeddings extracted by doc2vec~\cite{doc2vec} instead of using the texts in the ontology. CMF is originally designed for binary classification of a event happening or not. We replace their task head with a typical ConvTransE decoder to enable link prediction. 
    
    \item \textbf{Graph disentanglement methods} aim to separate the intertwined relational information into disentangled representations. We take into account two representative graph disentanglement methods: \textbf{DisenGCN}~\cite{DisenGCN} and \textbf{DisenKGAT}~\cite{DisenKGAT}. Both of the methods are designed for static graphs. 
    %To the best of our knowledge, graph disentanglement has not yet been introduced to the temporal knowledge graph.
\end{itemize}

\subsubsection{Hyper-parameter Settings}
We implement all the static methods using OpenKE~\footnote{https://github.com/thunlp/OpenKE}, for TANGO, RE-NET, RE-GCN, and EvoKG, we use their released code. For CMF and HiSMatch, we re-implement them by ourselves since these methods have not released the code. To be fair and following previous settings~\cite{REGCN}, for all the baselines and our method, we set $d=200$, use cross-entropy loss, search learning rate from \{0.01, 0.001, 0.0001\} and weight decay from $\{10^{-4}, 10^{-5}, 10^{-6}, 10^{-7}\}$. For temporal methods, we search the historical graph length $D$ in the range of \{1,3,7\}. For our method, we search the number of RGCN propagation layers $L$ from \{1,2,3\}, the number of hypergraph propagation layers $P$ from \{1,2\}, the number of LDA topics (\aka contexts) $K$ from \{3, 5, 7\}. We use Adam~\cite{Adam} optimizer and Xavier~\cite{Xavier} initialization for all the parameters. 

\begin{table*}[t]
\caption{The overall performance comparison between SeCoGD and baselines.}
\vspace{-0.1in}
\label{tab:overall_performance}
\centering
\setlength{\tabcolsep}{1mm}{
    \resizebox{0.95\textwidth}{!}{
        \begin{tabular}{l | cccc | cccc | cccc}
        \hline
        \multirow{2}{*}{Model} & \multicolumn{4}{c|}{EG} &\multicolumn{4}{c|}{IR} &\multicolumn{4}{c}{IS} \\
        \cline{2-13} & MRR & HIT@1 & HIT@3 & HIT@10 & MRR & HIT@1 & HIT@3 & HIT@10 & MRR & HIT@1 & HIT@3 & HIT@10 \\
        \hline
        \hline
        \textbf{DistMult}~\cite{DistMult}     & 0.1164 & 0.0344 & 0.1214 & 0.2927 & 0.1349 & 0.0392 & 0.1468 & 0.3379 & 0.1031 & 0.0223 & 0.0929 & 0.2950 \\
        \textbf{ConvE}~\cite{ConvE}           & 0.1151 & 0.0312 & 0.1272 & 0.2882 & 0.1365 & 0.0409 & 0.1485 & 0.3400 & 0.1060 & 0.0251 & 0.0984 & 0.2935 \\ 
        \textbf{ConvTransE}~\cite{ConvTransE} & 0.1205 & \underline{0.0377} & 0.1305 & 0.2921 & 0.1405 & 0.0462 & 0.1529 & 0.3412 & 0.1079 & 0.0287 & 0.0994 & 0.2930 \\
        \textbf{RotatE}~\cite{RotatE}         & 0.0892 & 0.0125 & 0.0772 & 0.2748 & 0.1055 & 0.0125 & 0.1074 & 0.3152 & 0.0879 & 0.0132 & 0.0714 & 0.2638 \\
        \textbf{RGCN}~\cite{RGCN}             & 0.0974 & 0.0279 & 0.1046 & 0.2377 & 0.1185 & 0.0366 & 0.1301 & 0.2860 & 0.0861 & 0.0242 & 0.0652 & 0.2307 \\
        \hline
        \hline
        \textbf{TANGO}~\cite{TANGO}  & 0.1043 & 0.0240 & 0.1106 & 0.2761 & 0.1249 & 0.0281 & 0.1367 & 0.3314 & 0.0972 & 0.0171 & 0.0852 & 0.2889 \\
        %\textbf{xERTE}~\cite{xERTE}  & & & & & & & & & & & & \\
        \textbf{RE-NET}~\cite{RENET} & 0.1212 & 0.0413 & 0.1224 & 0.2932 & 0.1401 & 0.0451 & 0.1501 & 0.3452 & 0.1064 & 0.0263 & 0.1016 & 0.2894 \\
        \textbf{RE-GCN}~\cite{REGCN} & \underline{0.1245} & 0.0352 & \underline{0.1366} & \underline{0.3101} & \underline{0.1647} & \underline{0.0622} & \underline{0.1796} & \underline{0.3838} & \underline{0.1301} & 0.0408 & \underline{0.1281} & \underline{0.3346} \\
        \textbf{EvoKG}~\cite{EvoKG}  & 0.0797 & 0.0012 & 0.0775 & 0.2529 & 0.0892 & 0.0011 & 0.0767 & 0.3120 & 0.0779 & 0.0008 & 0.0518 & 0.2789 \\
        \textbf{HiSMatch}~\cite{HiSMatch} & 0.1126 & 0.0275 & 0.1279 & 0.2906 & 0.1469 & 0.0496 & 0.1599 & 0.3572 & 0.1283 & \underline{0.0434} & 0.1248 & 0.3017 \\
        \hline
        \hline
        %\textbf{Glean}~\cite{Glean}     & & & & & & & & & & & & \\
        $\textbf{CMF}_{ont}$~\cite{CMF} & 0.1206 & 0.0348 & 0.1298 & 0.3015 & 0.1527 & 0.0529 & 0.1643 & 0.3673 & 0.1248 & 0.0368 & 0.1224 & 0.3256 \\
        $\textbf{CMF}_{art}$~\cite{CMF} & 0.1202 & 0.0345 & 0.1293 & 0.3027 & 0.1510 & 0.0496 & 0.1636 & 0.3716 & 0.1263 & 0.0382 & 0.1236 & 0.3261 \\
        \hline
        \hline
        \textbf{DisenGCN}~\cite{DisenGCN} & 0.0849 & 0.0196 & 0.0805 & 0.2198 & 0.1084 & 0.0275 & 0.1096 & 0.2793 & 0.0833 & 0.0162 & 0.0633 & 0.2427  \\
        %\textbf{DisenHAN}~\cite{DisenHAN} & & & & & & & & & & & & \\
        \textbf{DisenKGAT}~\cite{DisenKGAT} & 0.0801 & 0.0083 & 0.0822 & 0.2382 & 0.0895 & 0.0059 & 0.0977 & 0.2744 & 0.0724 & 0.0106 & 0.0429 & 0.2322 \\
        \hline
        \hline 			 						
        \textbf{SeCoGD(ours)} & \textbf{0.1464} & \textbf{0.0593} & \textbf{0.1605} & \textbf{0.3236} & \textbf{0.1757} & \textbf{0.0724} & \textbf{0.1902} & \textbf{0.3975} & \textbf{0.1552} & \textbf{0.0595} & \textbf{0.1588} & \textbf{0.3693} \\				
        %\textbf{\%Improv.}  & 17.65 & 57.29 & 17.49 & 4.34 & 6.68 & 16.32 & 5.48 & 3.55 & 19.24 & 45.90 & 23.98 & 10.36 \\
        \textbf{\%Improv.}  & 17.59 & 57.29 & 17.50 & 4.35 & 6.68 & 16.40 & 5.90 & 3.57 & 19.29 & 37.10 & 23.97 & 10.37 \\
        \hline
        \end{tabular}
    }
}
\vspace{-0.1in}
\end{table*}   

\subsection{Performance Comparison (RQ1)} \label{subsec:performance_comparison}
 Table~\ref{tab:overall_performance} shows the overall performance of our model and baselines. First of all, our method outperforms all the baselines on all three datasets. Among all the metrics, the improvement on HIT@1 is the highest, which is truly helpful in practice. Second, for all the baselines, RE-GCN achieves the best performance and even beats the models with textual inputs (\ie $\text{CMF}_{ont}$ and $\text{CMF}_{art}$), demonstrating its superiority in modeling temporal event graphs. This is why we select RE-GCN for context-specific modeling in our implementation. Third, in terms of the methods with textual inputs, $\text{CMF}_{ont}$ and $\text{CMF}_{art}$ perform well, beating most of the static and temporal methods. The results imply that the additional textual information offers valuable clues that are crucial to forecast future events. However, they are not the strongest baseline, probably because they are originally designed for binary event classification and the link prediction head is not perfectly adapted. Finally, for the two graph disentanglement-based methods, \ie DisenGCN and DisenKGAT, they do not perform very well. There are two possible reasons: 1) they rely on the static global graph, which cannot model the temporal evolving patterns; and 2) more importantly, the events in current datasets are coarse-grained and less discriminative, therefore, the methods that solely rely on structured data fail to learn disentangled representations. The results also justify our method that leverages the context as a prior guidance, instead of graph structure, to separate the event graph.

\subsection{Study of Key Modules (RQ2)} \label{subsec:model_study}
We conduct model studies to analyze the effect of the key modules in the two stages, \ie separation and collaboration. 

\begin{figure}[t]
    \centering
    \includegraphics[width = 0.95\linewidth]{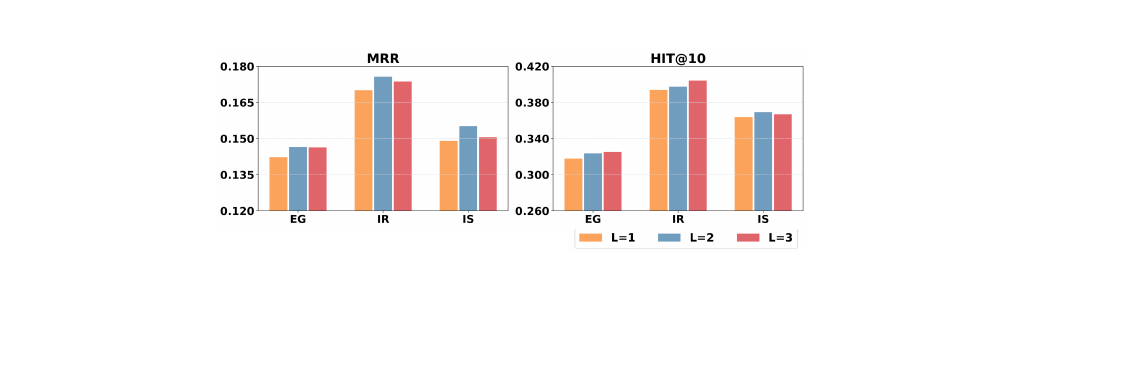}
    \vspace{-12pt}
    \caption{Results comparison of propagating different number of layers L in the context-specific modeling module.}
    \label{fig:RGCN_layer}
    \vspace{-8pt}
\end{figure}

\begin{figure}[t]
    \centering
    \includegraphics[width = 0.95\linewidth]{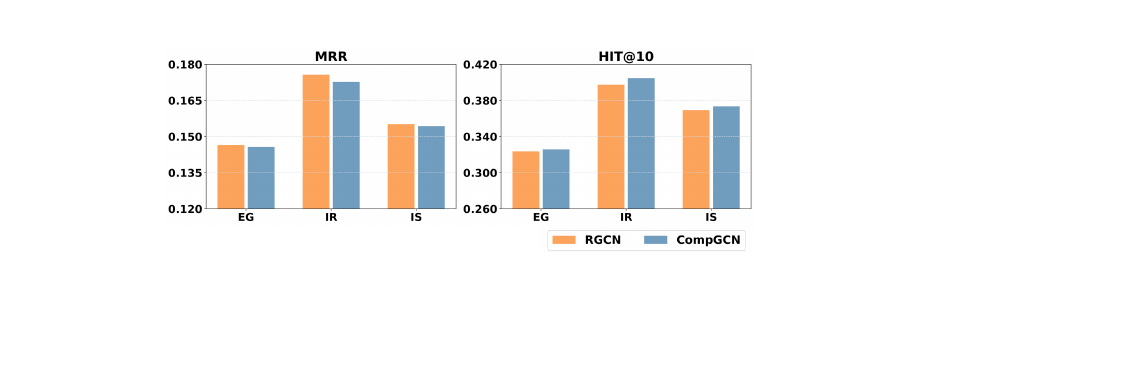}
    \vspace{-12pt}
    \caption{Results of using different graph kernels.}
    \label{fig:kernel}
    \vspace{-8pt}
\end{figure}

\subsubsection{Study of the Separation Stage}
%We adopt the RE-GCN backbone to model the concurrent and temporal patterns in each context. 
For the \textbf{concurrent event modeling}, we use the RGCN kernel. We tune the number of propagation layers, and the results are shown in Figure~\ref{fig:RGCN_layer}. Basically, two and three layers are better than one layer, depicting that higher-order information propagation over the concurrent event graph is beneficial to capture the context-specific signals. We also replace RGCN with CompGCN~\cite{CompGCN}, and Figure~\ref{fig:kernel} illustrates the results. Overall speaking, CompGCN and RGCN perform similarly to each other on the three datasets, and they differ slightly in terms of different evaluation metrics. It shows that our framework is not sensitive to relational modeling models.

\begin{figure}[t]
    \centering
    \includegraphics[width = 0.95\linewidth]{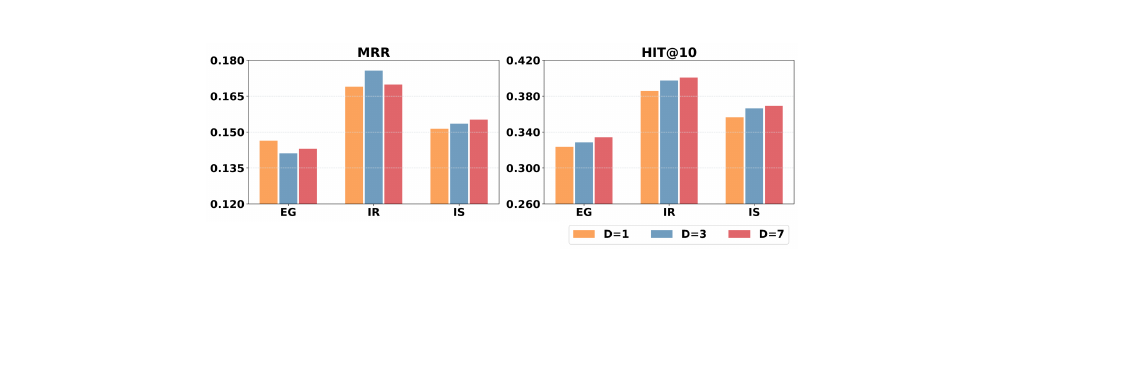}
    \vspace{-12pt}
    \caption{Results of with different historical length D.}
    \label{fig:hislen}
    \vspace{-8pt}
\end{figure}

For the \textbf{temporal pattern modeling} module, we tune the length of historical graphs $D$ that we used to generate the entity and relation embeddings. We try different historical length $D$ within \{1, 3, 7\} and visualize the results in Figure~\ref{fig:hislen}. For HIT@10, the longer the historical length is, the better the performance will be. But for MRR, EG and IR achieve the best performance with D=1 and D=2, respectively. This difference reminds practitioners to properly select evaluation metrics according to the application scenarios. For example on the EG dataset, if we care more about the ranking of the prediction, we need to choose MRR and set D=1. Meanwhile, if we pay more attention to the hit rate of the top-10 predicted results, HIT@10 with D=7 should be a better option. In addition, longer historical length takes extra computational costs. Therefore, it is a trade-off between efficacy and efficiency in practice.

\begin{table}[t]
\begin{center}
%\vspace{-0.1in}
\caption{Study of the cross-context modeling and context-aware prediction.}
\vspace{-0.1in}
\label{tab:model_study}
    \resizebox{0.45\textwidth}{!}{
        \begin{tabular}{c | cc | cc | cc}
            \hline
            \multirow{2}{*}{Model} & \multicolumn{2}{c|}{EG} & \multicolumn{2}{c|}{IR} & \multicolumn{2}{c}{IS} \\
            \cline{2-7}
             & MRR & H@10 & MRR & H@10 & MRR & H@10  \\
            \hline
            \hline
            \textbf{SeCoGD} & 0.146 & 0.324 & 0.176 & 0.397 & 0.155 & 0.369 \\
            \hline
            \hline
            \textbf{w/o Ent HG} & 0.139 & 0.315 & 0.168 & 0.391 & 0.147 & 0.359 \\
            \textbf{w/o Rel HG} & 0.143 & 0.331 & 0.170 & 0.400 & 0.147 & 0.362 \\
            \textbf{w/o Ent or Rel HG} & 0.138 & 0.315 & 0.163 & 0.386 & 0.144 & 0.355 \\
            \hline
            \hline
            \textbf{Avr. Context} & 0.130 & 0.309 & 0.163 & 0.373 & 0.129 & 0.331 \\
            \hline
            %\hline
            %\textbf{RE-GCN} & 0.125 & 0.310 & 0.165 & 0.384 & 0.130 & 0.335 \\
            %\hline
        \end{tabular}
    }
\end{center}
\vspace{-0.15in}
\end{table}

\subsubsection{Study of the Collaboration Stage}
We construct a hypergraph over the sub-embeddings of each entity and relation to retain the collaborative associations across multiple contexts. To test the efficacy of the collaboration stage and the implementation of the hypergraph, we design several ablated models by progressively removing the two hypergraphs of entity and relation. In Table~\ref{tab:model_study}, "w/o Ent HG", "w/o Rel HG", and "w/o Ent or Rel HG" refer to \textit{without relation hypergraph}, \textit{without entity hypergraph}, and \textit{without either entity or relation hypergraph}, respectively. From the results, we can see that the results of removing either relation or entity hypergraph are worse than SeCoGD but better than that of removing both, demonstrating the efficacy of both hypergraphs. More interestingly, the performance drop of removing the entity hypergraph is generally larger than that of removing the relation hypergraph, implying that the collaboration of entities is more valuable.

During prediction, our context-aware event forecasting will pair each query $(s,r,t+1)$ with an auxiliary context $c$. By specifying the context, its corresponding branch of the decoder will be selected and performs the forecasting. We argue that such a context-aware prediction narrows down the candidate space and performs better. To justify our hypothesis, we curate a variant, in which we do not specify the context during inference while just averaging the prediction scores from all the context decoders, corresponding to the row "Avr. Context" in Table~\ref{tab:model_study}. We can observe that "Avr. Context" performs much worse than SeCoGD. This phenomenon indicates that the specification of the proper context during inference is crucial to SeCoGD, justifying our hypothesis that the context plays a pivotal role in accurate event forecasting.

\subsection{Study of the Context (RQ3)} \label{subsec:context_study}
%As the main contribution of this work, context is the key to the overall task and solution. 
%We specifically investigate several characteristics of the context.

\begin{figure}[t]
    \centering
    \includegraphics[width = 0.95\linewidth]{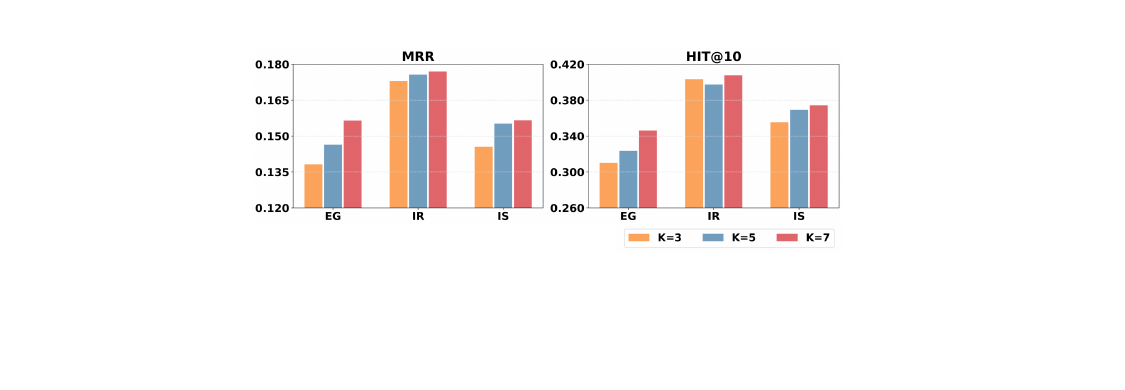}
    \vspace{-12pt}
    \caption{Results with different number of contexts.}
    \label{fig:context}
    \vspace{-8pt}
\end{figure}

\subsubsection{Effect of the Number of Contexts.}
We vary the number of LDA topics $K$ when we generate the context based on the news article, resulting in multiple versions of datasets with different number of contexts. We implement SeCoGD on all the versions of dataset and obtain the results, which are presented in Figure~\ref{fig:context}. In general, more contexts yield better performance. This is natural and reasonable because when the number of contexts increases, each context will be more specific, resulting in more fine-grained information being injected into the event. However, more contexts inevitably introduce more computational expenses. We leave the study of efficiency and scalability improvement in future work.

\begin{table}[t]
\begin{center}
%\vspace{-0.1in}
\caption{Alternative context generation methods.}
\vspace{-0.1in}
\label{tab:context_study}
    \resizebox{0.45\textwidth}{!}{
        \begin{tabular}{c | cc | cc | cc}
            \hline
            \multirow{2}{*}{Model} & \multicolumn{2}{c|}{EG} & \multicolumn{2}{c|}{IR} & \multicolumn{2}{c}{IS} \\
            \cline{2-7}
             & MRR & H@10 & MRR & H@10 & MRR & H@10  \\
            \hline
            \hline
            \textbf{RE-GCN} & 0.125 & 0.310 & 0.165 & 0.384 & 0.130 & 0.335 \\
            \hline
            \hline
            \textbf{K-means} & 0.139 & 0.314 & 0.169 & 0.388 & 0.145 & 0.352 \\
            \textbf{GMM} & 0.139 & 0.316 & 0.165 & 0.375 & 0.134 & 0.339 \\
            \hline
            \hline
            \textbf{LDA(SeCoGD)} & 0.146 & 0.324 & 0.176 & 0.397 & 0.155 & 0.369 \\
            \hline
        \end{tabular}
    }
\end{center}
\vspace{-0.15in}
\end{table}

\subsubsection{Effect of the Context Curation Methods.}
We define the context as a categorical label for each event, while LDA is just one of the automatic methods in order to avoid extensive labor and costs for context annotation. We argue that alternative automatic approaches are also workable for our framework. To illustrate this property, we leverage two prominent text clustering methods, \ie K-means and GMM (Gaussian Mixture Model) using the article embeddings pre-trained by doc2vec~\cite{doc2vec}, to generate contexts. Results based on the newly-generated contexts are shown in Table~\ref{tab:context_study}. From the results, we can conclude that SeCoGD is generally able to outperform RE-GCN by leveraging the contexts generated with alternative clustering methods. This further illustrates that our method is robust to diverse context sources. Nonetheless, our proposal of using LDA to generate contexts performs best, thus we take it as the default setting.

\begin{figure*}
    \centering
    \includegraphics[width = 0.98\linewidth]{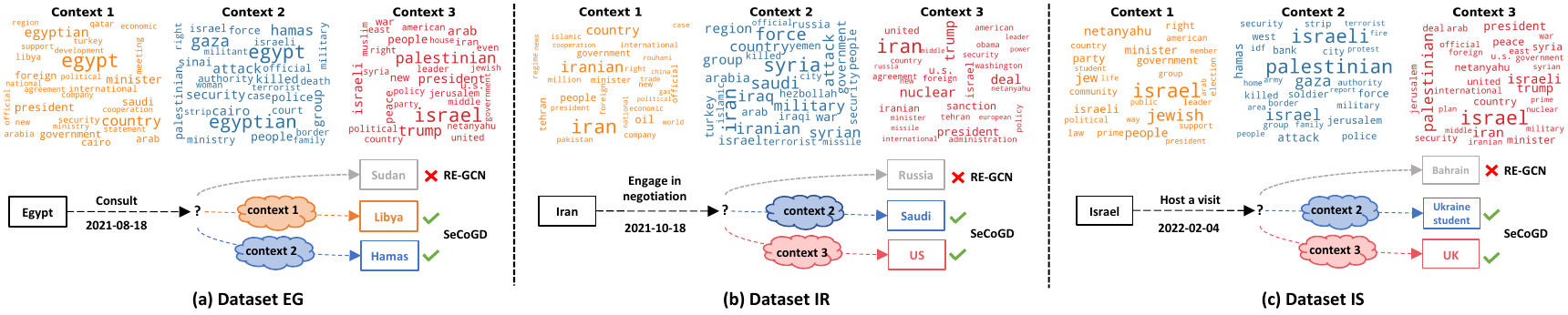}
    \vspace{-0.1in}
    \caption{Case study on three datasets. In each sub-figure, the context number K is set as three, the top shows the word cloud of each context, and the bottom illustrates several exemplar forecasting results by SeCoGD and RE-GCN.}
    \vspace{-0.15in}
    \label{fig:word_cloud}
\end{figure*}

\subsubsection{Case Study}

%\begin{figure}[t]
%    \centering
%    \includegraphics[width = 0.9\linewidth, height=0.7\textwidth]{figures/figure-case-study.pdf}
%    \vspace{-10pt}
%    \caption{Case study of context-aware event forecasting.}
%    \label{fig:word_cloud}
%    \vspace{-5pt}
%\end{figure}

%In addition to the quantitative evaluation, we also aim to qualitatively comprehend how the context contributes to event forecasting. In particular, 

We seek to elicit the content of each context, thus to elaborate how different events are predicted under distinct contexts. As shown in Figure~\ref{fig:word_cloud}, for each dataset, we illustrate the top words of each context in the form of word cloud~\cite{WordCloud} (the size of each word is proportional to its weight in the LDA topic distribution). From word clouds, we observe rich information within each context and clear content differences among contexts. For example, each context in the EG dataset covers background information such as popular actors, important cities, and critical actions; meanwhile, they are about economic, military, and political events respectively. 

We also pick an example query $(s, r, t+1)$ from the testing set of each dataset to concretely explicate the benefits of context-aware event forecasting. We list the object with the highest score predicted by RE-GCN and SeCoGD, and we observe that SeCoGD generates more accurate prediction results compared to RE-GCN. With a given context $c$, general event types such as `Consult', `Negotiate', and `Host a visit' are now narrated with more supplementary information modeled in the context, leading to better results. We also observe that given the same query, SeCoGD sometimes predicts distinct objects under different contexts. For example, for the example query in IS dataset, SeCoGD predicts that \textit{Israel} will host a visit for \textit{Ukraine students} under \textit{Context 2}, and predicts \textit{UK} instead under \textit{Context 3}. We notice that both events exist in the dataset and thus are both correct, and the two predictions are in line with the contents of their context. As shown in the word cloud for IS Context 2 and 3, the former prediction might take more military factors into consideration, and the latter is more related to government affairs. This demonstrates the flexibility in depicting the event by context.

\section{Related Work} \label{sec:related_work}
%We review the related works of this paper from two aspects: 1) temporal event forecasting, and 2) graph disentanglement.

\subsection{Temporal Event Forecasting}
Temporal event forecasting aims to forecast future events based on a list of observed historical events. It has been studied in various application scenarios, including criminal activities~\cite{crime}, disease outbreaks~\cite{disease}, stock markets~\cite{stock}, as well as international political events~\cite{gdelt,icews}. Various problem formulations are utilized with regard to different event types, such as time series forecasting, natural language generation, and link prediction. In this work, we follow the typical formulations of link prediction, which is also called temporal knowledge graph completion. It inherits from static knowledge graph completion, where the key is to learn relational embeddings via various scoring functions, such as TransE~\cite{TransE}, DistMult~\cite{DistMult}, ComplEx~\cite{complex}, RotatE~\cite{RotatE}, ConvE~\cite{ConvE}, ConvTransE~\cite{ConvTransE} \textit{etc.} To tackle the temporal evolving patterns and forecast future events, recurrent neural network (RNN)~\cite{LSTM} has been included. RE-NET~\cite{RENET} proposes to use RGCN~\cite{RGCN} to capture the relational patterns in each timestamp and GRU~\cite{GRU} to model the dynamics of embeddings over time. RE-GCN~\cite{REGCN} additionally incorporates a static graph to learn the static properties of the entities and adopts ConvTransE~\cite{ConvTransE} as the decoder. TANGO~\cite{TANGO} models the structure of candidate entities via neural ordinary differential equations; %xERTER~\cite{xERTE} leverages a dynamic pruning procedure to identify a relevant subgraph of the query; 
EvoKG~\cite{EvoKG} considers the time information for event forecasting; HiSMatch~\cite{HiSMatch} reformulates event forecasting as a query-candidate matching problem and proposes a two-branch framework to match a query to candidate entities. 
%CENET~\cite{CENET} uses contrastive learning to contrast seen and unseen entities. 
%CRNet~\cite{CRNet} models the concurrent structures in the query time. 
More related works~\cite{TTransE,HIP,TITer,CyGNet,xERTE} can be seen in the survey~\cite{TKG-survey}.
%Even though great success has been achieved in the TKG modeling for event forecasting, 
Most of these TKG methods only operate on pure structured data, overlooking the rich semantic or contextual information. To address these limitations, Glean~\cite{Glean} and CMF~\cite{CMF} propose to use the textual information. They simplify the event forecasting problem from fine-grained link prediction to an easier binary classification problem, \ie predicting whether an event will happen or not. In addition, the textual information is only available for historical events but unavailable for future events, thus this additional information cannot directly narrow down the candidate space. Some works also use context for event prediction, while they are either for event status classification~\cite{eventCLScontext} or time series forecasting~\cite{timeseriesContext}, which are different tasks.
%In summary, all of the current event forecasting methods have not considered the categorical contexts.

\subsection{Graph Disentanglement}
Graph neural networks~\cite{GCN,GAT,GraphSage} have been the defacto solutions for graph representation learning. Graph disentanglement is the extension of disentangled representation learning from the general domain to the graph data. Disentangled representation learning focuses on separating the unified representation into multiple disentangled components, thus achieving many excellent modeling properties such as enhanced representation capability or explainability. Various studies have been conducted on CV~\cite{image-disen}, NLP~\cite{text-disen}, as well as recommender system~\cite{MacridVAE,DGCF,GNUD}. For graph representation learning, disentanglement has also garnered particular attention. DisenGCN~\cite{DisenGCN} is one of the pioneering works to use multiple disentangled graph convolutional kernels to learn disentangled node representations. FactorGCN~\cite{FactorGCN} factorizes the node embedding into multiple blocks, which captures interpretable global topological semantics. IPGDN~\cite{IPGDN} leverages the Hilbert-Schmidt Independence Criterion (HSIC) to achieve disentanglement. ADGCN~\cite{ADGCN} introduces adversarial learning to graph disentanglement representation learning. DisenHAN~\cite{DisenHAN} is designed for heterogeneous graph, where multiple node and relation types are involved. The most relevant work for event forecasting is DisenKGAT~\cite{DisenKGAT}, which aims to learn disentangled representations for knowledge graph.
Despite various studies on graph disentanglement, our work differs from current works and promotes these works in several aspects. First, we are the first to introduce graph disentanglement learning to temporal event forecasting. Second, most of the existing works aim to directly disentangle the graph purely using the graph's own features, ignoring the contextual information. %Third, current works that solely rely on graph features target at enlarging the representation distance of disentangled components with each other, however, our prior-guided disentanglement naturally endows our model sufficient discriminative capability. 
%Therefore, a key challenge for our approach is how to capture the collaborative association between the disentangled representations.
\section{Conclusion and Future Work} \label{conclusion}

In this work, we explored the incorporation of context into the problem of event forecasting and proposed a novel task of context-aware event forecasting. To tackle this novel problem, we borrowed the idea from graph disentanglement and designed an overall framework SeCoGD. Specifically, we utilized the context as prior guidance to separate the event graph and incorporated a context-specific modeling module to capture the relational and temporal patterns in each context. In addition, we designed a cross-context modeling module to model the collaborative associations among multiple contexts. Since there are no available datasets for this new task, we built three large-scale datasets based on GDELT. Extensive experiments on these three datasets demonstrated that our framework outperforms all the SOTA methods. Various model studies further elaborated more details about the effectiveness of the key modules and various contexts of the framework.

Despite the progress achieved by this work, there are several limitations, thus motivating multiple potential research directions in the future. First, the implementation of context generation is based on unsupervised methods, while human-generated contexts, such as tags and categories, could be more useful in practice. Second, the original articles of these events are just used as a proxy to generate contexts, of which just a little information has been utilized. More effective approaches to mining more beneficial patterns from raw texts are promising. 
%Third, SeCoGD just makes use of the predefined contexts to separate the event graph and uses the hypergraph to collaborate among contexts, which may be sub-optimal. 
Third, more advanced graph disentanglement methods are expected to be explored and enhance the performance.
Finally, in addition to next step prediction, the more important yet challenging multi-horizon forecasting should be studied in future.

\section*{acknowledgement}
This research is supported by the Defence Science and Technology Agency. This research is also supported by the National Natural Science Foundation of China (9227010114) and the University Synergy Innovation Program of Anhui Province (GXXT-2022-040).

%\newpage
\bibliographystyle{ACM-Reference-Format}
\bibliography{0_main}
\end{document}